\documentclass{article}[12pt]
\usepackage{amsmath}
\usepackage{cite,times}
\numberwithin{equation}{section}

\def\pr{{\rm pr}}
\def\rd{{\rm d}}
\def\re{{\rm e}}
\def\rank{{\rm rank}}

\begin{document}
\title{Lie Symmetries and Exact Solutions of First Order Difference 
Schemes\footnote{To be published in Journal of Physics A: Mathematical and General, 2004
}}
\author{M. A. Rodr\'{\i}guez\\ 
Dept. F\'{\i}sica Te\'orica II, 
Facultad de F\'{\i}sicas\\ 
Universidad Complutense, 
20840-Madrid, Spain\\
 and\\
P. Winternitz\\
Centre de recherches math\'ematiques et \\
D\'epartement de Math\'ematiques et Statistique,
Universit\'e de Montr\'eal\\
CP 6128, Succ Centre-Ville,
Montr\'eal (Qu\'ebec)
H3C 3J7 Canada}
\date{}

\maketitle
\begin{abstract}
We show that any first order ordinary differential equation with a 
known Lie point symmetry group can be discretized into a difference 
scheme with the same symmetry group. In general, the lattices are not 
regular ones, but must be adapted to the symmetries considered. The 
invariant difference schemes can be so chosen that their solutions 
coincide exactly with those of the original differential equation.
\end{abstract}

\section{Introduction}

The purpose of this article is to analyze Lie point symmetries and to
obtain exact solutions of first order difference schemes.  These are
two-point schemes of the form:
\begin{eqnarray}\label{eq}
&&E_{a}(x,y,x_{+},y_{+})=0,\quad a=1,2, \\
&&\left|\frac{\partial(E_{1},E_{2})}{\partial (x_{+},y_{+})}\right|
\neq 0,
\end{eqnarray}
where we use the notation $x\equiv x_{n}$, $x_{+}\equiv x_{n+1}$,
$y\equiv y_{n}$, $y_{+}\equiv y_{n+1}$ for the independent and
dependent variables, evaluated at two different points.  The two
equations (\ref{eq}) define a difference equation, as well as a
lattice.  The general solution of the scheme depends on two constants
and has the form
\begin{equation}\label{sol}
   y=y(n,C_{1},C_{2}),\quad x=x(n,C_{1},C_{2}).
\end{equation}
  
Equation (\ref{sol}) can also be rewritten as:
 \begin{equation}\label{soldos}
   y=y(x,C_{1},C_{2}),\quad x=x(n,C_{1},C_{2}).
\end{equation}
The formula for $y$ can be interpreted as interpolating between the
points of the lattice and determining the solution $y$ for all values
of $x$.

A standard equally spaced lattice is given by a specific choice of one
of the two equations (\ref{eq}), namely $E_{2}=x_{+}-x-h=0$, where $h$
is a constant (the lattice spacing) and its solution is $x_{n}=n h+
x_{0}$ (where $x_{0}$ is one of the two integration constants of the
scheme).

Instead of the variables $x$, $y$, $x_{+}$, $y_{+}$, we can use
\begin{equation}\label{alt}
    x,\quad y,\quad h=x_{+}-x,\quad y_{x}=\frac{y_{+}-y}{x_{+}-x}.
\end{equation}
The continuous limit $h\to 0$ of the scheme (\ref{eq}) is then obvious,
namely, one of the equations should turn into a first order ordinary
differential equation (ODE), the other into an identity (like $0=0$).
We shall write the ODE as
\begin{equation}\label{ode}
E(x,y,y')=y'-F(x,y)=0.   
\end{equation}
    
The symmetries and solutions of the difference scheme (\ref{eq}) that
we shall obtain below will in the limit $h\to 0$ turn into Lie point
symmetries and solutions of the ODE (\ref{ode}).  To establish the
connection, let us recall some well known results on symmetries of
first order ODEs \cite{Ol79}.

The Lie point symmetry group of the ODE (\ref{ode}) is always
infinite-dimensional \cite{Ol79}.  Its Lie algebra, the ``symmetry algebra'', is
realized by vector fields of the form
\begin{equation}\label{vec}
    X=\xi(x,y)\partial_{x}+\phi(x,y)\partial_{y}
\end{equation}
satisfying
\begin{equation}\label{pr}
    \pr X (E)\big|_{E=0}=0.
\end{equation}
In (\ref{pr}) $\pr X$ is the first prolongation of $X$,
i.e., \cite{Ol79}
\begin{eqnarray}\label{prolog}
    \pr  X&=&\xi(x,y)\partial_{x}+ 
    \phi(x,y)\partial_{y}+\phi^{x}(x,y,y')\partial_{y'},\\
    \phi^{x}&=&\phi_{x}+(\phi_{y}-\xi_{x})y'-\xi_{y}(y')^2,\nonumber
\end{eqnarray}    
where the subscripts are partial derivatives. 
Equation (\ref{pr}) amounts to a single first order linear partial differential
equation for the two functions $\xi$ and $\phi$ and as such it has infinitely
many solutions. This is the reason why the Lie point symmetry of equation (\ref{ode})
is infinite-dimensional \cite{Ol79}. In general it may be difficult or
impossible to find any explicit analytical solution; as difficult as
finding an integrating multiplier. However, if we find at least one
particular explicit solution of equation  (\ref{pr}), we can obtain a one-dimensional
subalgebra of the symmetry algebra of equation (\ref{ode}). This is sufficient to
integrate equation (\ref{ode}) in quadratures.

All elementary methods of solving first order ODEs amount to special
cases of the above procedure.

A different application of the vector field $X$ and its
prolongation (\ref{prolog}) is to construct first order ODEs that are
invariant under a given Lie group of local point transformations,
namely, those generated by the vector field $X$.  In this case the
functions $\xi(x,y)$, $\phi(x,y)$ and hence also $\phi^{x}(x,y,y')$,
are known.  The invariant equation is obtained by solving the first
order partial differential equation:
\begin{equation}\label{partial}
[\xi(x,y)\partial_{x}+\phi(x,y)\partial_{y}+ \phi^x(x,y,y')
\partial_{y'}]E(x,y,y')=0    
\end{equation}
for the function $E(x,y,y')$.  Solving by the method of
characteristics, we obtain two elementary invariants:
\begin{equation}\label{inv}
    I_{1}=I_{1}(x,y),\quad I_{2}=I_{2}(x,y,y'),\quad \frac{\partial 
    I_{2}}{\partial y'}\neq 0.
\end{equation}
An invariant equation is given by any relation between $I_{1}$ and 
$I_{2}$, i.e.,
\begin{equation}\label{equat}
    E(I_{1},I_{2})=0,\quad \frac{\partial E}{\partial I_{2}}\neq 0.
\end{equation}    

If we are given a two-dimensional Lie algebra of vector fields
$\{X_{1},X_{2}\}$ and we require invariance under the two-dimensional
Lie group they generate, then two different possibilities can occur.
The first is that the two equations (\ref{partial}) (for $X_{1}$ and
$X_{2}$ respectively) have a common solution
\begin{equation}\label{invar}
    I=I(x,y,y'),\quad \frac{\partial I}{\partial y'}\neq 0.
\end{equation}
The invariant ODE then is
\begin{equation}\label{equati}
    F(I)=0,
\end{equation}
where $F$ is arbitrary.  We say that equation (\ref{equati}) is
``strongly invariant'' with respect to the group generated by $\{X_{1},
X_{2}\}$.  If no such invariant $I(x,y,y')$ exists, then we look for an
invariant manifold and a ``weakly invariant'' equation.  This is
obtained from the condition that the two equations $\pr X_{1}(E)=0$, $\pr
X_{2}(E)=0$ should be equivalent.  This is a condition on the matrix of
coefficients, i.e.,
\begin{equation}\label{ran}
    \rank\left(\begin{array}{ccc}
    \xi_{1} & \phi_{1} & \phi_{1}^{x}\\
    \xi_{2} & \phi_{2} & \phi_{2}^{x}
    \end{array}\right)=1.
\end{equation}
This condition, together with, say, $\pr X_{1} (E)=0$, provides the weakly
invariant equation.

In Section \ref{liepoint} we will adapt the above results to the case of the 
difference system (\ref{eq}) and in Sections 3-9 consider many 
examples. The examples will be difference analogs of ODEs with known 
symmetry groups (linear equations, separable equations, etc).

In each case we first present a one- or two-dimensional symmetry algebra
of the ODE and use it to solve the equation. We have not found these
symmetry algebras in the literature, but they are implicit in standard
integration procedures (we have made them explicit). In each case we spell
out the invariant difference schemes and choose one that has exactly the same
general solution as the ODE.

\section{Lie point symmetries and first order difference schemes}
\label{liepoint}

The point of view that we will be taking here is the same as in 
previous publications, e.g. 
\cite{Do91,Do01,LW91,LW02,LT00,LT01,Wi04,Ma80,QS93,DK00,DK04}
and references therein. 
Namely, Lie point symmetries of difference equations will be 
continuous point transformations $(x,y)\to (\tilde{x},\tilde{y})$, 
taking solutions of the system (\ref{eq}) into solutions. They will 
be induced by a Lie algebra of vector fields of the same form 
(\ref{vec}) as for differential equations. The prolongation of the 
vector field will be different. No derivatives figure in equation 
(\ref{eq}); instead we prolong to other points on the lattice. In the 
case of the system (\ref{eq}) we have
\begin{equation}\label{prd}
    \pr^{D}X=\xi(x,y)\partial_{x}+\phi(x,y)
    \partial_{y}+\xi(x_{+},y_{+})\partial_{x_{+}}+
    \phi(x_{+},y_{+})\partial_{y_{+}}.
\end{equation}    
The continuous limit (\ref{prolog}) is recovered by putting
\begin{equation}\label{cont}
    x_{+}=x+h,\quad y_{+}\equiv y(x_{+})=y(x)+hy'(x)+\cdots
\end{equation}
then, expanding into Taylor series
\begin{eqnarray}
    \xi(x_{+},y_{+})&=&
    \xi(x,y)+h\xi_{x}(x,y)+hy'\xi_{y}(x,y)+\cdots,\\
    \phi(x_{+},y_{+})&=&
    \phi(x,y)+h\phi_{x}(x,y)+hy'\phi_{y}(x,y)+\cdots\nonumber
\end{eqnarray}
and acting with $\pr^{D} X$ on a function of $x$, $y$, $h$ and $y_{x}$ (see 
equation (\ref{alt})). We obtain
\begin{eqnarray}
    \pr^{D}X(F(x,y,h,y_{x})) &=& \{\xi(x,y)\partial_{x}+ 
    \phi(x,y)\partial_{y}+
    [\phi_{x}+(\phi_{y}-\xi_{x})y_{x}\nonumber\\ &&
    -\xi_{y}(y_{x})^2]\partial_{y_{x}}\}
    F(x,y,h,y_{x})+O(h)
\end{eqnarray}

Thus we have
\begin{equation}
    \lim_{h\to 0}\pr^{D} X=\pr X.
\end{equation}    
as required.

Using the prescription
\begin{equation}
    \pr^{D} X (E_{a})\big|_{E_{1}=E_{2}=0} =0   
\end{equation}
we can find the symmetries of a given system (\ref{eq}). Instead, we 
shall start from an ODE and its known symmetries and construct the 
invariant difference scheme from the known vector fields.

For a one-dimensional symmetry algebra we find invariants by 
solving the partial differential equation :
\begin{equation}\label{pde}
    [\xi(x,y)\partial_{x}+\phi(x,y)\partial_{y}+\xi(x_{+},y_{+}) 
    \partial_{x_{+}}+\phi(x_{+},y_{+})\partial_{y_{+}}]
    F(x,y,x_{+},y_{+})=0
\end{equation}
by the method of characteristics. The elementary invariants are:
\begin{equation}\label{inva}
    I_{1}^{D}=I_{1}^{D}(x,y),\quad I_{2}^{D}= I_{2}^{D}
    (x_{+},y_{+}),\quad  I_{3}^{D}=I_{3}^{D}(x,y,h,y_{x}),\quad
    \frac{\partial I_{3}}{\partial y_{x}}\neq 0
\end{equation}
with $h$ and $y_{x}$ as in equation (\ref{alt}). Any two relations 
between the expressions (\ref{inva}) will give an invariant difference 
scheme, for instance
\begin{equation}\label{invdiff}
    I_{3}^{D}=F(I_{1}^{D}),\quad I_{1}^{D}=I_{2}^{D}
\end{equation}
with $F$ chosen to obtain the correct continuous limit (we will drop 
the superscript $D$ below).

If we have a two-dimensional symmetry algebra, we will obtain two 
invariants. If one of them is of the type $I_{3}$ in (\ref{inva}), 
we again obtain a (strongly) invariant difference scheme. If not, we 
must look for an invariant manifold, given in this case by the rank 
condition
\begin{equation}\label{randos}
    \rank\left(\begin{array}{cccc}
    \xi_{1}(x,y) & \phi_{1}(x,y) & \xi_{1}(x_{+},y_{+}) & 
    \phi_{1}(x_{+},y_{+}) \\
    \xi_{2}(x,y) & \phi_{2}(x,y) & \xi_{2}(x_{+},y_{+}) &
    \phi_{2}(x_{+},y_{+})
    \end{array}\right)=1.
\end{equation}

We will see below for specific examples that for each invariant ODE 
we obtain invariant difference schemes with the same invariance 
group and the same general solution. Indeed, our choice of the 
invariant difference schemes
\begin{equation}\label{difsch}
    E_{a}(I_{1},I_{2},I_{3})=0,\quad a=1,2
\end{equation}
will be guided by two considerations:
\begin{description}
    \item{ i)} To obtain the original ODE in the continuous limit
    \item{ii)} To obtain a difference scheme that has exactly the same 
    general solution as the original ODE (for any value of the lattice 
    spacing $h=x_{+}-x$).
\end{description}

\section{Linear equations}
Let us consider the first order linear inhomogeneous ODE
\begin{equation}\label{lin}
    y'=a(x)y+b(x).
\end{equation}
For convenience we redefine the given functions $a(x)$ and 
$b(x)$, putting $a(x)\equiv A'(x)$, $b(x)\equiv
B'(x)\re^{A(x)}$. 
Equation (\ref{lin}) and its general solution are then written as
\begin{eqnarray}\label{solbis}
    y'&=&A'(x)y+B'(x)\re^{A(x)},\\
    y(x)&=&(B(x)+k)\re^{A(x)},\label{solter}
\end{eqnarray}
where $k$ is the integration  constant and the primes indicate 
$x$-derivatives.

Equation (\ref{solbis}) has a two-dimensional Lie point symmetry group, 
generated by the vector fields
\begin{equation}\label{lie}
    X_{1}=\re^{A(x)}\partial_{y},\quad 
    X_{2}=\left[y-B(x)\re^{A(x)}\right]\partial_{y}.
\end{equation}

Now let us look for a difference scheme invariant under the group 
generated by the Lie algebra (\ref{lie}). The prolongations of $X_{1}$ 
and $X_{2}$ to the space $\{x,y,x_{+},y_{+}\}$ have only two invariants 
$x$ and $x_{+}$. However they do allow an invariant manifold, given by
the condition (\ref{randos}), which in this case reduces to
\begin{equation}\label{sche}
    y_{+}\re^{-A(x_{+})}-y\,\re^{-A(x)}-B(x_{+})+B(x)=0.
\end{equation}
Adding an invariant lattice equation, e.g.,
\begin{equation}\label{schem}
    x_{+}-x=h
\end{equation}
we obtain an invariant difference scheme (\ref{sche}), (\ref{schem}). 
Not only does this system reduce to the ODE (\ref{solbis}) in the
continuous limit, but equation (\ref{solter}) gives the exact general
solution of the discrete system (\ref{sche}), (\ref{schem}) (for any
value of $h$).

Notice that while equation  (\ref{sche}) is linear in $y$, it is not the
difference equation one would get by the usual numerical discretization. 
To compare the two, let us reintroduce the discrete variable $n$, 
putting $x=x_{n}$, $x_{+}=x_{n+1}$, $y=y_{n}$, 
$y_{+}=y_{n+1}$. 

A ``naive'' discretization would be:
\begin{equation}
    \frac{y_{n+1}-y_{n}}{x_{n+1}-x_{n}}=A'(x_{n}) 
    y_{n}+B'(x_{n})\re^{A(x_{n})}
\end{equation}
or possibly
\begin{equation}
    \frac{y_{n+1}-y_{n}}{x_{n+1}-x_{n}}=\frac{A(x_{n+1})-A(x_{n})}{x_{n+1}-x_{n}} 
    y_{n}+\frac{B(x_{n+1})-B(x_{n})}{x_{n+1}-x_{n}}\re^{A(x_{n})}
\end{equation}
with $x_{n}=nh$ in both cases. The system (\ref{sche}), 
(\ref{schem}), on the other hand, is rewritten as;
\begin{eqnarray}\label{syst}
y_{n+1}\re^{-A(x_{n+1})}-y_{n}\,\re^{-A(x_{n})}-B(x_{n+1})+B(x_{n})=0
\\    \nonumber
x_{n+1}-x_{n}=h
\end{eqnarray}

All of the above discretizations coincide in the limit $h\to 0$, 
however only the discretization (\ref{sche}), (\ref{schem}), i.e. 
(\ref{syst}), is exact in the sense that it has exactly the same 
general solution (\ref{solter}) as the ODE (\ref{lin}).

Similar comments hold for all the discretizations that we present 
below in Sections 4-9. We shall not repeat them each time.

\section{Separable equations}
Let us consider the separable ODE
\begin{equation}\label{sep}
    y'=f(x)g(y)
\end{equation}
and for convenience redefine $f(x)\equiv A'(x)$, $g(y)\equiv
1/\dot{B}(y)$ (the prime is an $x$-derivative, the dot a
$y$-derivative). The (implicit) general solution of equation (\ref{sep}) 
is
\begin{equation}\label{solsep}
    B(y)=A(x)+k,
\end{equation}
where $k$ is a constant. The equation itself is rewritten as:
\begin{equation}\label{newsep}
    y'=\frac{A'(x)}{\dot{B}(y)}.
\end{equation}
Equation (\ref{newsep}) has a two-dimensional symmetry group, 
generated by
\begin{equation}\label{alg}
    X_{1}=\frac{1}{\dot{B}(y)}\partial_{y},\quad 
    X_{2}=\frac{1}{A'(x)}\partial_{x}.\quad 
\end{equation}

The group invariants of the group generated by the algebra 
(\ref{alg}) in the discrete space are
\begin{equation}
    I_{1}=A(x_{+})-A(x),\quad I_{2}=B(y_{+})-B(y).
\end{equation}
Using $I_{1}$ and $I_{2}$, we write an invariant difference scheme as
\begin{eqnarray}\label{sch}
    B(y_{+})-B(y)-A(x_{+})+A(x) & = & 0,\\
    A(x_{+})-A(x) & = & \epsilon(h),
\end{eqnarray}
where $\epsilon$ is some constant, satisfying $\epsilon(h)\to 0$ for
$h\to 0$.  Equation (\ref{solsep}) clearly provides the general
solution of equation (\ref{sch}).

As a specific example, let us choose
$$    B(y)=y^M,\quad A(x)=x^N.$$
The difference scheme
\begin{eqnarray}
    y_{+}^M-y^M-x_{+}^N+x^N & = & 0,  \\
    x_{+}^N-x^N & = & \epsilon
\end{eqnarray}
is solved by
\begin{equation}\label{solu}
    y=(x^N+k)^{1/M},\quad x_{n}=(n\epsilon +x_{0}^N)^{1/N},
\end{equation}
where we choose $\alpha>0$, $x_{0}\ge 0$. Clearly $y(x)$ as in 
(\ref{solu}) also solves the ODE obtained in the continuous limit, 
namely
\begin{equation}
    y'=\frac{Nx^{N-1}}{My^{M-1}}.
\end{equation}

\section{Exact equations}

We will consider in this section exact equations, that is, equations 
of the form:
\begin{equation}\label{exac}
    y'(x)=-\frac{A(x,y)}{B(x,y)},\quad {\rm i.e.,}\quad
    A(x,y)\rd x+B(x,y)\rd y=0
\end{equation}
satisfying
\begin{equation}
    A_{y}=B_{x},\quad {\rm i.e.,}\quad A(x,y)=V_{x}(x,y),\; 
    B(x,y)=V_{y}(x,y)
\end{equation}
for some function $V(x,y)$.

Equation (\ref{exac}) is invariant under a one-dimensional group 
generated by
\begin{equation}\label{vecexac}
    X=B(x,y)\partial_{x}-A(x,y)\partial_{y}.
\end{equation}    
The general solution of (\ref{exac}) is given implicitly by the 
relation
\begin{equation}\label{solexac}
    V(x,y)=k,
\end{equation}
where $k$ is an integration constant.

In the discrete case equation (\ref{pde}) leads to the characteristic 
system
\begin{equation}
    \frac{\rd x}{V_{y}(x,y)}=-\frac{\rd y}{V_{x}(x,y)}= \frac{\rd 
    x_{+}}{V_{y_{+}}(x_{+},y_{+})}=
    -\frac{\rd y_{+}}{V_{x_{+}}(x_{+},y_{+})}
\end{equation}
and hence to three invariants
\begin{eqnarray}
    I_{1}=V(x,y),\quad I_{2}=V(x_{+},y_{+})\label{int}\\
    I_{3}=\int\frac{\rd x_{+}}{V_{y_{+}}(x_{+},y_{+}(x_{+},I_{2}))}-
    \int\frac{\rd x}{V_{y}(x,y(x,I_{1}))}.\label{intdos}
\end{eqnarray}
To obtain the integrals involved in $I_{3}$ we have solved the 
equations (\ref{int}) for $y$ and $y_{+}$. The 
discrete version of equation (\ref{exac}) is:
\begin{equation}\label{discexac}
    V(x_{+},y_{+})-V(x,y)=0
\end{equation}    
with (\ref{solexac}) as its solution.  Equation (\ref{intdos}) can be
used to define the invariant lattice.

As a specific example, consider the ODE
\begin{equation}
      y'=\frac{1}{2(x+y)}-1.
\end{equation}
It is exact and we have
\begin{equation}
      V=(x+y)^2-x.
\end{equation}    
The vector field (5.3) in this case is
\begin{equation}
      X=2(x+y)\partial_{x}-(2(x+y)-1)\partial_{y}.
\end{equation}
The invariants in the discrete case are
\begin{eqnarray}
      &I_{1}=(x+y)^2-x,\quad I_{2}=(x_{+}+y_{+})^2-x_{+},&\\
      &I_{3}=x_{+}+y_{+}-x-y.&
    \nonumber
\end{eqnarray}   
The invariant difference scheme can be written as
\begin{eqnarray}
      &(x_{+}+y_{+})^2-(x+y)^2 =x_{+}-x,&\\
      &x_{+}+y_{+}-x-y=\epsilon.&
\end{eqnarray}

Returning to the general case, we see that the discrete analogue of an
exact ODE is equation (\ref{discexac}).  The invariant lattice can be
given by equation (\ref{intdos}).  In the continuous limit, we have
$I_{3}\to 0$, i.e., $x_{+}\to x$ and (\ref{discexac}) goes to
\begin{equation}
    V_{x}+V_{y} y_{x}=0.
\end{equation}    

\section{Homogeneous equations}
Let us consider the first order ODE
\begin{equation}\label{hom}
    y'=x^{k-1}F\left(\frac{y}{x^k}\right),
\end{equation}
where $F$ is an arbitrary smooth function and $k$ is a real constant. 
This is the most general first order ODE invariant under the scaling 
group
\begin{equation}
    \tilde{y}=\re^{\lambda k}y,\quad \tilde{x}=\re^{\lambda}x
\end{equation}
generated by the vector field
\begin{equation}\label{vechom}
    X=x\partial_{x}+ky\partial_{y}.
\end{equation}
For convenience, we replace the function
$F(t)$ in equation (\ref{hom}) by $F(t)=1/\dot{H}(t)+kt$, so 
equation  (\ref{hom}) is rewritten as
\begin{equation}\label{homdos}
    y'=\frac{x^{k-1}}{\dot{H}\left(\frac{y}{x^k}\right)}+k\frac{y}{x}.
\end{equation}

The change of variables $(x,y(x))\to (t,z(t))$ with
\begin{equation}
    t=\frac{y}{x^k},\quad z=\log x,\qquad x=\re^{z},\quad y=t\re^{kz}
\end{equation}
will straighten out the vector field $X$ and transform equation 
(\ref{hom}) into
\begin{equation}\label{homeq}
    z_{t}=\dot{H}(t).
\end{equation}
Solving equation (\ref{homeq}) and returning to the original variables,
we obtain the general solution of equation (\ref{hom}) in the form
\begin{equation}\label{solhom}
    y(x)=x^k H^{-1}(\log x-C),
\end{equation}
where $C$ is an integration constant and $H^{-1}$ is the function 
inverse to $H(t)$.

Let us now find the invariant difference scheme corresponding to 
equation (\ref{hom}). Prolonging the vector field $X$ of 
equation (\ref{vechom}) as in equation (2.1), we find three elementary 
invariants
\begin{equation}
    I_{1}=\frac{y}{x^k},\quad I_{2}=\frac{y_{+}}{x_{+}^k},\quad 
    I_{3}=\frac{x_{+}}{x}.
\end{equation}    

Using them we write an invariant difference scheme as:
\begin{eqnarray}
    \log I_{3}-H(I_{2})+H(I_{1}) &=& 0,\\
    I_{3}-1-\epsilon &=& 0.
\end{eqnarray}

More explicitly, we have
\begin{eqnarray}\label{schom}
    \log x_{+}-\log x-H\left(\frac{y_{+}}{x_{+}^k}\right)+
    H\left(\frac{y}{x^k}\right) &=& 0,\\
    x_{+}-x-\epsilon x &=& 0,\label{schomdos}
\end{eqnarray}
where $\epsilon$ is some constant.  The $\epsilon\to 0$ limit of the
difference scheme (\ref{schom}), (\ref{schomdos}) is the ODE
(\ref{homdos}), as required, and its general solution is
(\ref{solhom}), together with
\begin{equation}\label{homlatt}
    x_{n}=(\epsilon+1)^n x_{0}.
\end{equation}
As in the previous examples, the invariant difference scheme has the 
same exact solution as the original ODE.

As a specific example, we take $k=1$ and the ODE
\begin{equation}\label{homex}
    y'=\frac{x^2+y^2}{xy}.
\end{equation}
The difference scheme in this case is
\begin{equation}\label{homexdos}
    \left(\frac{y_{+}}{x_{+}}\right)^2-
    \left(\frac{y}{x}\right)^2=2\log
    \frac{x_{+}}{x},\quad \frac{x_{+}-x}{x}=\epsilon.
\end{equation}
The solution of both (\ref{homex}) and (\ref{homexdos}) is
\begin{equation}
    y=x\sqrt{2\log x-C}
\end{equation}
(together with (\ref{homlatt}) in the discrete case).

\section{Rotationally invariant equations}

Let us consider a rotation in the $(x,y)$ space. It is generated by
\begin{equation}\label{vecrot}
    X=y\partial_{x}-x\partial_{y}.
\end{equation}    
The most general first order ODE invariant under these rotations is:
\begin{equation}\label{rot}
    y'=\frac{K(\rho)y-x}{y+K(\rho)x},\quad \rho=\sqrt{x^2+y^2}.
\end{equation}
To solve equation (\ref{rot}) we straighten out the vector field 
(\ref{vecrot}) by going to polar coordinates $(x,y(x))\to (\rho, 
\alpha(\rho))$
\begin{equation}\label{polar}
    x=\rho\cos \alpha,\quad y=\rho\sin\alpha.
\end{equation}
Equation (\ref{rot}) reduces to
\begin{equation}
\alpha_{\rho}=-\frac{1}{\rho K(\rho)}
\end{equation}    
and we obtain
\begin{equation}
\alpha(\rho)=-\int\frac{1}{\rho K(\rho)}\rd\rho
\end{equation}    

For simplicity, let us restrict to the special case $K(\rho)=K=$ constant. 
We then obtain the solution of (\ref{rot}) as a logarithmic spiral
\begin{equation}\label{spi}
    K\alpha+\log \rho =\log \rho_{0}
\end{equation}
or, in the original variables
\begin{equation}\label{rotsol}
    \frac{1}{2}\log (x^2+y^2)+K \arctan \frac{y}{x}=\log \rho_{0}.
\end{equation}

Now let us consider the discrete case. The prolonged vector field is:
\begin{equation}
    \pr X =y\partial_{x} -x\partial_{y}+y_{+}\partial_{x_{+}} 
    -x_{+}\partial_{y_{+}}.
\end{equation}
This means that the pairs $(x,y)$ and $(x_{+},y_{+})$ will transform 
like vectors undergoing a rotation in a Euclidean plane. We can form 
4 invariants
\begin{equation}
    I_{1}=x^2+y^2,\quad I_{2}=x_{+}^2+y_{+}^2,\quad 
    I_{3}=xy_{+}-x_{+}y,\quad I_{4}=xx_{+}+yy_{+}
\end{equation}    
with one relation between them, namely
\begin{equation}
    I_{3}^2+I_{4}^2=I_{1}I_{2}.
\end{equation}

As an invariant difference scheme, we write
\begin{eqnarray}\label{rotsch}
    E_{1} & = & \log\frac{I_{2}}{I_{1}}+2K\arctan 
    \frac{I_{3}}{I_{4}} =0,\\ \label{rotschdos}
    E_{2} & = & \frac{1}{2}(I_{2}-I_{1})+K I_{3}=0
\end{eqnarray}
with $K$ a constant. Using equation (\ref{cont}) we can check that the 
continuous limit of both these expressions is the ODE (\ref{rot}) 
(with $K$ constant). In general we have
\begin{equation}
    \left|\frac{\partial(E_{1},E_{2})}{\partial(x_{+},y_{+})}
    \right|_{h\neq 0}\neq 0,\quad \lim_{h\to 0}
    \left|\frac{\partial(E_{1},E_{2})}{\partial(x_{+},y_{+})}
    \right|=0.
\end{equation}

Expression (\ref{rotsol}), the exact solution of the ODE (\ref{rot}), is 
also an exact solution of the system (\ref{rotsch}), (\ref{rotschdos}).

The situation is more transparent in polar coordinates (\ref{polar}). For 
$K=$ constant, the ODE is
\begin{equation}\label{dif}
    \frac{\rd\alpha}{\rd \rho}=-\frac{1}{K\rho}.
\end{equation}
The prolongation of the vector field corresponding to rotational 
invariance in the discrete case is
\begin{equation}
    \pr X=\partial_{\alpha}+\partial_{\alpha_{+}}.
\end{equation}

The invariants are
\begin{equation}
    I_{1}=\alpha_{+}-\alpha,\quad I_{2}=\rho_{+},\quad I_{3}=\rho.
\end{equation}
An invariant difference scheme corresponding to the ODE (\ref{dif}) is
\begin{eqnarray}
    \alpha_{+}-\alpha=-\frac{1}{K}(\log 
    \rho_{+}-\log \rho),\label{rotdis}\\
    \rho_{+}-\rho=\epsilon.\label{rotdisdos}
\end{eqnarray}    
The exact solution of (\ref{dif}) and (\ref{rotdis}) is given by
equation (\ref{spi}).  The lattice determined by equation
(\ref{rotdisdos}) is uniform:
\begin{equation}
    \rho_{n}=\epsilon n+\rho_{0}.
\end{equation}

\section{Invariant difference schemes on uniform lattices}

Let us consider the ODE
\begin{equation}\label{equni}
    y'=-\frac{A_{x}}{A_{y}}+f'(x)\frac{1}{A_{y}},\quad A_{y}\neq 0,
\end{equation}
where $A(x,y)$ and $f(x)$ are some smooth functions. Equation 
(\ref{equni}) is invariant under transformations generated by
\begin{equation}\label{vecuni}
    X=\frac{1}{A_{y}(x,y)}\partial_{y}.
\end{equation}    
As a matter of fact, any first order ODE invariant under 
transformations generated by a vector field of the form
\begin{equation}
    X=\phi(x,y)\partial_{y}
\end{equation}    
can be reduced to the form (\ref{equni}).

The general solution of equation  (\ref{equni}) can be written 
implicitly as
\begin{equation}\label{soluni}
    A(x,y)=f(x)+C.
\end{equation}

In the discrete case the first prolongation of the vector field 
(\ref{vecuni}) has 3 invariants; they can be chosen to be:
\begin{equation}
    I_{1}=x_{+},\quad I_{2}=x,\quad I_{3}=A(x_{+},y_{+})- A(x,y).
\end{equation}
An invariant difference scheme on a uniform lattice, having the ODE 
as a continuous limit is:
\begin{eqnarray}\label{invdif}
    A(x_{+},y_{+})-A(x,y)=f(x_{+})-f(x),\\
    x_{+}-x=h.
\end{eqnarray}
Its solution is given by equation (\ref{soluni}) together with
\begin{equation}\label{hlatt}
    x_{n}=hn+x_{0}.
\end{equation}

As a specific example, let us choose
\begin{equation}
    X=xy\partial_{y}.
\end{equation}
The invariant ODE in this case is
\begin{equation}\label{eqcont}
    y'=\frac{y}{x}\log y+q'(x)xy
\end{equation}    
the corresponding difference scheme is
\begin{equation}\label{eqdisc}
    \frac{1}{x_{+}}\log y_{+}-\frac{1}{x}\log y -q(x_{+}) 
    +q(x)=0,\quad x_{+}-x=h.
\end{equation}
The solution of both (\ref{eqcont}) and (\ref{eqdisc}) is
\begin{equation}
    y=\re^{x(C+q(x))},
\end{equation}
where $C$ is an integration constant.

The vector field (\ref{vecuni}) is not the only one compatible with a 
uniform lattice. Another one is
\begin{equation}
X=\partial_{x}+\phi(x,y)\partial_{y}
\end{equation}
for any function $\phi(x,y)$. As an example, let us take
\begin{equation}\label{vecunidos}
    X=\partial_{x}+x^a y^b\partial_{y}.
\end{equation}
The corresponding invariant first order ODE is
\begin{equation}\label{eqinv}
    y'=(k(\zeta)+x^a)y^b,\quad \zeta=\frac{1}{a+1}x^{a+1} 
    +\frac{1}{b-1}y^{-b+1},
\end{equation}    
where $k$ is any function of $\zeta$. For $k=k_{0}=$constant. The 
solution of equation (\ref{eqinv}) is
\begin{equation}\label{eqinvsol}
    y=(1-b)^{\frac{1}{1-b}}\left[k_{0}x+\frac{1}{a+1} 
    x^{a-1}+C\right]^{\frac{1}{1-b}},
\end{equation}
where $C$ is an integration constant.

The invariants of the discrete prolongation of the vector field 
(\ref{vecunidos}) are
\begin{equation}
    I_{1}=x_{+}-x,\quad I_{2}=\frac{x^{a+1}}{a+1}+\frac{y^{-b+1}}
    {b-1},\quad 
    I_{3}=\frac{x_{+}^{a+1}}{a+1}+\frac{y_{+}^{-b+1}}{b-1}.
\end{equation}
We write an invariant difference scheme as
\begin{equation}\label{invsch}
    I_{2}-I_{3}-k_{0}I_{1}=0,\quad I_{1}=h
\end{equation}    
with $k_{0}$ and $h$ constant. More explicitly (\ref{invsch}) is
\begin{eqnarray}\label{expl}
    \frac{1}{1-b}[y^{1-b}_{+}-y^{1-b}]-\frac{1}{1+a}[ x_{+}^{a+1}- 
    x^{a+1}]-k_{0}(x_{+}-x)=0,\\
    x_{+}-x=h.
\end{eqnarray}    
The continuous limit of (\ref{expl}) is the ODE (\ref{eqinv}).
Moreover, equation (\ref{eqinvsol}) is the exact solution of
(\ref{expl}) for any value of $h$ (not just the limit $h\to 0$).

\section{Invariant difference schemes on exponential lattices}

The vector field (\ref{vecuni}) is also compatible with an 
invariant scheme of the form (\ref{invdif}) on the exponential lattice
\begin{equation}\label{explat}
    \frac{x_{+}-x}{x}=\epsilon,\quad {\rm i.e.}\quad
    x_{n}=(\epsilon+1)^n x_{0}
\end{equation}
with $\epsilon=$ constant.  Taking the limit $\epsilon\to 0$ we again
obtain the ODE (\ref{equni}).

Another symmetry, leaving the lattice (\ref{explat}) invariant, is 
generated by the vector field
\begin{equation}
    X=x\partial_{x}+\phi(x,y)\partial_{y}
\end{equation}
with $\phi(x,y)$ arbitrary.

As an example, consider
\begin{equation}
    X=x\partial_{x}+x^a y^b\partial_{y},\quad a\neq 0,\; b\neq 1.
\end{equation}
The corresponding invariant ODE is
\begin{equation}\label{expinv}
    y'=k(\zeta)\frac{1}{x}y^b+x^{a-1}y^b,\quad \zeta=\frac{1}{a} x^a+ 
    \frac{1}{b-1}y^{1-b}.
\end{equation}
For $k=k_{0}$ constant, the solution is
\begin{equation}\label{expsol}
    y=(1-b)^{\frac{1}{1-b}}\left[k_{0}\log x+ 
    \frac{x^a}{a}+C\right]^{\frac{1}{1-b}}.
\end{equation}

In the discrete case, the prolongation of $X$ has three 
invariants
\begin{equation}
    I_{1}=\frac{x_{+}}{x},\quad I_{2}=\frac{y^{1-b}}{b-1} 
    +\frac{x^a}{a},\quad 
    I_{3}=\frac{y_{+}^{1-b}}{b-1}+\frac{x_{+}^a}{a}.
\end{equation}
An invariant scheme having (\ref{expinv}) with $k=k_{0}$ as its limit 
and (\ref{expsol}) as its solution is
\begin{equation}
    I_{2}-I_{3}-k_{0}\log I_{1}=0,\quad I_{1}-1=\epsilon,
\end{equation}
i.e.,
\begin{eqnarray}
    \frac{y_{+}^{1-b}}{1-b}-\frac{y^{1-b}}{1-b}-\frac{x_{+}^a}{a}+ 
    \frac{x^a}{a} -k_{0}(\log x_{+}-\log x)=0,\\
    x_{+}-x=\epsilon x_{+}.
\end{eqnarray}

\section{Conclusions}

Looking directly for symmetries of a difference scheme of the type 
(\ref{eq}) is not a particular fruitful enterprise. Just as in the 
case of first order ODEs one gets an underdetermined system of equations. 
Infinitely many solutions exist, but there is no algorithm for 
finding them.

We have taken the complementary point of view.  We have postulated the
form of a vector field, then found ODEs and difference schemes,
invariant under the corresponding symmetry group.  The symmetry makes
it possible to solve the ODE exactly analytically.  The invariant
differential scheme then has the same solution.  More precisely, the
symmetry leads to a family of difference schemes, one of which has
solutions coinciding with those of its continuous limit (the original
ODE).

Essentially we have constructed a partial ``catalogue'' of exactly
solvable two-point schemes.  This corresponds to a list of exactly
solvable first order ODEs: linear equations, separable equations,
exact equations, homogeneous equations, etc.  The ``complete list'' is
infinite: for any chosen realization of a one- or two-dimensional Lie
algebra we can construct an invariant ODE and an invariant difference
scheme.

Without the symmetries to guide us, the obtained difference scheme are 
not obvious at all. Starting from an ODE (\ref{ode}) and discretizing 
in a standard way, i.e., replacing the derivative $\rd y/\rd x$ by 
a discrete derivative and writing
\begin{equation}\label{deriv}
    \frac{y_{n+1}-y_{n}}{x_{n+1}-x_{n}}=F(x_{n},y_{n}),\quad x_{n}=h n 
    +x_{0}
\end{equation}
we would loose virtually all symmetries. Moreover, the exact solution 
of equation (\ref{deriv}) would differ from that of the ODE 
(\ref{ode}) by terms of the order $h$. For our symmetry dictated 
difference schemes, the exact solutions coincide with those of the 
ODEs. The main result of this article can be summed up as follows.

{\bf Theorem.}
For every first order ODE there exists an invariant two point difference scheme with exactly the same general solution as the ODE.

{\it Proof.} Consider the ODE (\ref{ode}) and assume that we know its general solution in the form of a first integral
\begin{equation}\label{solt}
h(x,y)=K,\quad h_y\neq 0.
\end{equation}
The general element (\ref{vec}) of the symmetry algebra of equation (\ref{ode}) must annihilate the function $h(x,y)$. The equation $Xh=0$ implies that $X$ has the form:
\begin{equation}\label{vect}
X=\xi(x,y)\left(\partial_x-\frac{h_x}{h_y}\partial_y\right),
\end{equation}
where $\xi(x,y)$ is an arbitrary smooth function.

Let us now find the invariants of the group action induced by the vector field (\ref{vect}) in the space $(x,x_+,y,y_+)$. They are obtained by solving the characteristic system:
\begin{equation}
\frac{\rd x}{\xi(x,y)}=\frac{\rd x_+}{\xi(x_+,y_+)}=
\frac{h_y(x,y)\rd y}{-\xi(x,y)h_x(x,y)}=
\frac{h_{y_+}(x_+,y_+)\rd y_+}{-\xi(x_+,y_+)h_{x_+}(x_+,y_+)}.
\end{equation}
The invariants hence are
\begin{eqnarray}
&& I_1=h(x,y),\quad I_2=h(x_+,y_+),\nonumber\\
&& I_3=\int\frac{\rd x}{\xi(x,y(x,I_1))}-\int\frac{\rd x_+} {\xi(x_+,y_+(x_+,I_2))}\;,
\end{eqnarray}
where the function $\xi(x,y)$ can be freely chosen.

A difference scheme that has (\ref{solt}) as its general solution is
\begin{equation}
I_1=I_2,\quad I_3=c,\quad  (c={\rm constant}).
\end{equation}
\hfill QED

Three comments are in order here. 
\begin{enumerate}
\item The proof given above is not a constructive one. It assumes that we already know the general solution (\ref{solt}) of the ODE. This was not assumed in the rest of this article, where we constructed the difference schemes using only one, or sometimes two elements of the symmetry algebra of an ODE.

\item The arbitrariness in the function $\xi(x,y)$ can be put to good use in the choice of lattices. For instance, choosing $\xi(x,y)=1$ we obtain a uniform lattice as in (\ref{hlatt}). Choosing $\xi=x$ we obtain an exponential lattice, as in equation (\ref{explat}).

\item The theorem and the results of this article are specific of first order ODEs and two point difference schemes.
\end{enumerate}

Let us compare with the case of second order ODEs and three-point 
difference schemes. S. Lie classified all (complex) second order ODEs 
into equivalence classes according to their Lie point symmetries 
\cite{Li24}. A similar classification of three-point difference 
schemes is much more recent \cite{DK00}. It was shown \cite{DK04} that if a 
three-point difference scheme has a symmetry group of dimension 3 (or 
larger) with at least a two-dimensional subalgebra of Lagrangian 
symmetries, then the scheme can be integrated analytically. A 
crucial element in the integration was the existence of a Lagrangian 
and the interpretation of the difference scheme as a discrete 
analogue of an Euler-Lagrange equation \cite{DK00,DK04}.

The fact that we obtain differential equations and difference schemes 
that have identical symmetries and solutions has interesting 
implications. It suggests a certain duality between continuous and 
discrete descriptions of physical phenomena. Thus, exactly the same 
physical predictions may be described by some continuous curve, or 
by a series of points on this curve, distributed with an arbitrary 
density.

\section*{Acknowledgements}
We thank V.A. Dorodnitsyn, D. Levi and P. Morando for interesting
discussions.  The research of P.W. was partly supported by research
grants from NSERC of Canada and FQRNT du Qu\'ebec.  The research of
M.A.R. was partly supported by a research grant from MCyT of Spain
(BFM2002-02646).  Both authors profited from the NATO collaborative
research grant PST.CLG978431.

\end{document}